\documentclass{bmvc2k}
\usepackage{multirow}
\usepackage{bbm}
\usepackage{amsfonts}

\title{Exemplar Learning for Medical Image Segmentation}

\addauthor{Qing En}{QingEn@cunet.carleton.ca}{1}
\addauthor{Yuhong Guo}{YuhongGuo@carleton.ca}{1,2}

\addinstitution{
 Carleton University\\
 Ottawa, Canada
}
\addinstitution{
 Canada CIFAR AI Chair\\
 Amii,
 Canada
}

\runninghead{EN, GUO}{Exemplar Learning for Medical Image Segmentation}

\def\eg{\emph{e.g}\bmvaOneDot}

\def\etal{\emph{et al}\bmvaOneDot}
\def\ie{\emph{i.e}\bmvaOneDot}
\begin{document}

\maketitle

\begin{abstract}
Medical image annotation typically requires expert knowledge and hence incurs time-consuming and expensive data annotation costs. To alleviate this burden, we propose a novel learning scenario, Exemplar Learning (EL), to explore automated learning processes for medical image segmentation with a single annotated image example. 
This innovative learning task is particularly suitable for medical image segmentation, where all categories of organs can 
be presented in one single image and annotated all at once. 
To address this challenging EL task, we propose an Exemplar Learning-based Synthesis Net (ELSNet) framework for medical image segmentation that enables innovative exemplar-based data synthesis, pixel-prototype based contrastive embedding learning, and pseudo-label based exploitation of the unlabeled data. Specifically, ELSNet introduces two new modules for image segmentation:  
an exemplar-guided synthesis module, which enriches and diversifies the training set by synthesizing annotated samples from the given exemplar,
and a pixel-prototype based contrastive embedding module, which enhances the discriminative capacity of the base segmentation model via contrastive representation learning. 
Moreover, we deploy a two-stage process for segmentation model training,
which exploits the unlabeled data with predicted pseudo segmentation labels.  
To evaluate this new learning framework, 
we conduct extensive experiments on several organ segmentation datasets and present an in-depth analysis. 
The empirical results show that the proposed exemplar learning framework produces
effective segmentation results. 
\end{abstract}

%-------------------------------------------------------------------------
\section{Introduction}
\label{sec:intro}
Medical image analysis is becoming increasingly important for clinical diagnosis and surgical planning due to the rapid advancement of medical imaging technologies \cite{duncan2000medical,tajbakhsh2016convolutional}.
Notably, medical image segmentation is one of the critical steps in quantitative medical image analysis, aiming to automatically identify the target region from medical images pixel-by-pixel \cite{sharma2010automated,milletari2016v,hesamian2019deep}. 
Fully supervised deep neural networks have been demonstrated to yield desirable segmentation results by using large amounts of labeled training data \cite{chen2021transunet,wang2022mixed}.
However, obtaining abundant annotated medical images at the pixel-level entails substantial labour and financial expenses because annotating medical images
requires the knowledge of clinical experts that is not always available.
To reduce the annotation cost, several techniques have been developed to perform medical image segmentation with less annotated data \cite{fei2006one,choudhury2021unsupervised}, including semi-supervised segmentation \cite{mittal2019semi,reiss2021every,wu2021collaborative} and few-shot segmentation \cite{sung2018learning,roy2020squeeze}.
\begin{figure*}
\centering
  \includegraphics[width=0.75\linewidth,height=40mm]{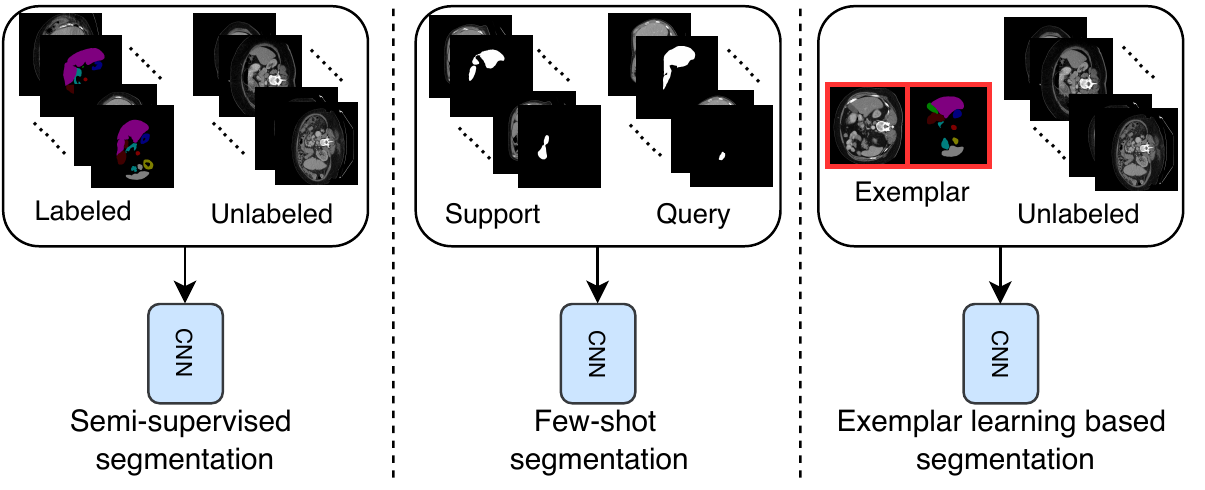}\\
\caption{
Exemplar learning vs. semi-supervised segmentation and few-shot segmentation.
In semi-supervised segmentation, multiple labeled samples and many unlabeled samples are present.
In few-shot segmentation, many support and query samples are available from the base categories.
By contrast, only one labeled sample and many unlabeled samples are available in 
exemplar learning. 
}
\label{fig:fig1}
\vskip -.2in
\end{figure*}

Although some notable improvements have been achieved, current solutions are still unable to eliminate the labelling conundrum.
Most few-shot segmentation methods rely on extensive auxiliary training datasets with exhaustive annotated data to transfer knowledge from the support set to the query set \cite{roy2020squeeze,tang2021recurrent}.
Semi-supervised segmentation methods usually focus on exploiting the consistency property of unlabeled data, but they still require a nontrivial portion of the densely annotated data \cite{wu2021collaborative,seibold2021reference,luo2021semi}.
We observe that the images for medical segmentation tasks often contain variations of the same set of organ categories, while a proper example image can cover all the parts for the whole organ category set. 
Motivated by this observation, we propose a novel learning scenario called Exemplar Learning (EL) to set up the working environment for a new set of medical image segmentation techniques that require only one single expert annotated image.
The differences between this new EL setting and the previous semi-supervised segmentation and the few-shot segmentation settings are illustrated in Figure \ref{fig:fig1}. 

The fundamental challenges for exemplar learning lie in the following two aspects:
(1) Data diversity is severely deficient, and the number of foreground-background pixels is imbalanced in medical image datasets.
Since only one annotated image is available, the model can easily be overfitted to the labeled sample.
(2) The contrast level between the organ and the background is low, and the differences among multiple organs' appearances are barely discernible, easily resulting in the lack of 
discriminative
capacity for the segmentation models.
This phenomenon makes it difficult to distinguish the boundaries between organs, 
leading to over-segmentation issues.
The fact that humans can learn by analogy \cite{bar2007proactive,ryali2021learning,zhao2021distilling} motivates us to address the abovementioned challenges by enriching the sample diversity and the discriminability of models.

In this paper, we propose a novel framework, \textit{ELSNet}, for learning to segment medical images effectively with only one annotated image.
ELSNet enriches the diversity of the labeled data by synthesizing training data and enhances the discriminability of the base segmentation model by performing pixel-prototype based contrastive embedding learning.
Specifically, given the exemplar image with all organ categories labeled, we first devise an exemplar-guided synthesis module (ESM) to enlarge our training set by taking crops of foreground organs and pasting them through various transformations onto different background images.
This can increase the invariance of organ representations to different backgrounds, while enriching the diversity of the sample.
Next, we design a pixel-prototype based contrastive embedding module (PCEM) to decompose the organs into distinct and consistent parts by capturing homogeneous components of the same type through contrastive embedding learning.
This module enables pixels belonging to the same organ to be similar, in contrast to the case where pixels belong to different organs, and hence is expected to improve the discriminability of the segmentation model.
Moreover, we deploy a two-stage process for segmentation model training, which exploits the unlabeled data with predicted pseudo segmentation labels
to further improve the segmentation model.
The main contributions of our paper can be summarized as follows:
\begin{itemize}
\setlength\itemsep{0em}		
\item 
We propose a novel learning scenario, Exemplar Learning, which 
investigates medical image segmentation with a single annotated image.
\item
We propose a novel ELSNet framework to segment medical images in the EL scenario by creating exemplar-based synthetic data, learning pixel-prototype based contrastive embeddings,	and exploiting unlabeled data with pseudo-labels.
\item
Experimental results on two medical image segmentation datasets show that the proposed ELSNet can effectively perform the medical semantic segmentation task.
\end{itemize}

\section{Related Work}

\paragraph{Semi-Supervised Medical Image Segmentation.}
Semi-supervised semantic segmentation has received increasing attention to train models by reducing the mask labeling cost \cite{souly2017semi,mittal2019semi,zou2020pseudoseg,ouali2020semi,sohn2020fixmatch}.
The technique has also been applied to the field of medical image segmentation \cite{reiss2021every,wu2021collaborative,seibold2021reference,luo2021semi}.
\citet{reiss2021every} 
proposed a multi-label deep supervision model to supervise low-resolution features and applied it to multiple medical supervision signals.
\citet{wu2021collaborative} presented a semi-supervised polyp segmentation model by collaborative and adversarial learning.
Moreover, \citet{seibold2021reference} used labeled images as references to generate more accurate pseudo-labels.
\citet{luo2021semi} proposed a dual-task-consistency semi-supervised framework for medical image segmentation. 
These semi-supervised methods require multiple annotated images for model training.
By contrast, we propose to train segmentation models with only one annotated image, which is more challenging.

\paragraph{Few-Shot Medical Image Segmentation.}
Few-shot segmentation has been exploited in the medical image domain \cite{zhao2019data,mondal2018few,ouyang2019data,yu2020foal,chen2020realistic}.
Most of these methods require many base categories to be annotated during the training phase and require fine-tuning for unseen classes.
SE-Net \cite{roy2020squeeze} and RP-Net \cite{tang2021recurrent} focus on designing models to segment unseen classes without retraining.
Sli2Vol \cite{yeung2021sli2vol} propagated the 2D image segmentation with an affinity matrix directly to reconstruct the rest of the image in 3D volumes in a self-supervised manner.
Ouyang \etal \cite{ouyang2020self} generated superpixel-based pseudo-labels and used the adaptive local prototype information for training the self-supervised FSS framework.
These methods require the support set in the test phase, and the predicted masks only contain foreground and background categories.
By contrast, the proposed exemplar learning does not require a support set, and the predicted masks have semantic information.

\section{Proposed Method}
In the setting of the exemplar learning, one labeled training image (\ie Exemplar) and T unlabeled training images are given, and denoted as $\mathcal{D}_{E} = (I_{e}, Y_{e})$ and $\mathcal{D}_{U} = \{(I_{u}^t)\}_{t=1}^{T}$.
An input image is defined as $I\in \mathbb{R}^{1*H*W}$, and the label is defined as $Y_{e}\in \{0, 1\}^{K*H*W}$, where $K, H, W$ are the number of categories in the dataset, height and width of the input image, respectively.
In this setting, the single exemplar image contains one segmentation instance for each category.
We propose an Exemplar Learning-based Synthesis Net (ELSNet) 
framework to train
a good segmentation model from the given input images. 
\begin{figure*}
\vskip .05in	
\centering
  \includegraphics[width=0.70\linewidth,height=73mm]{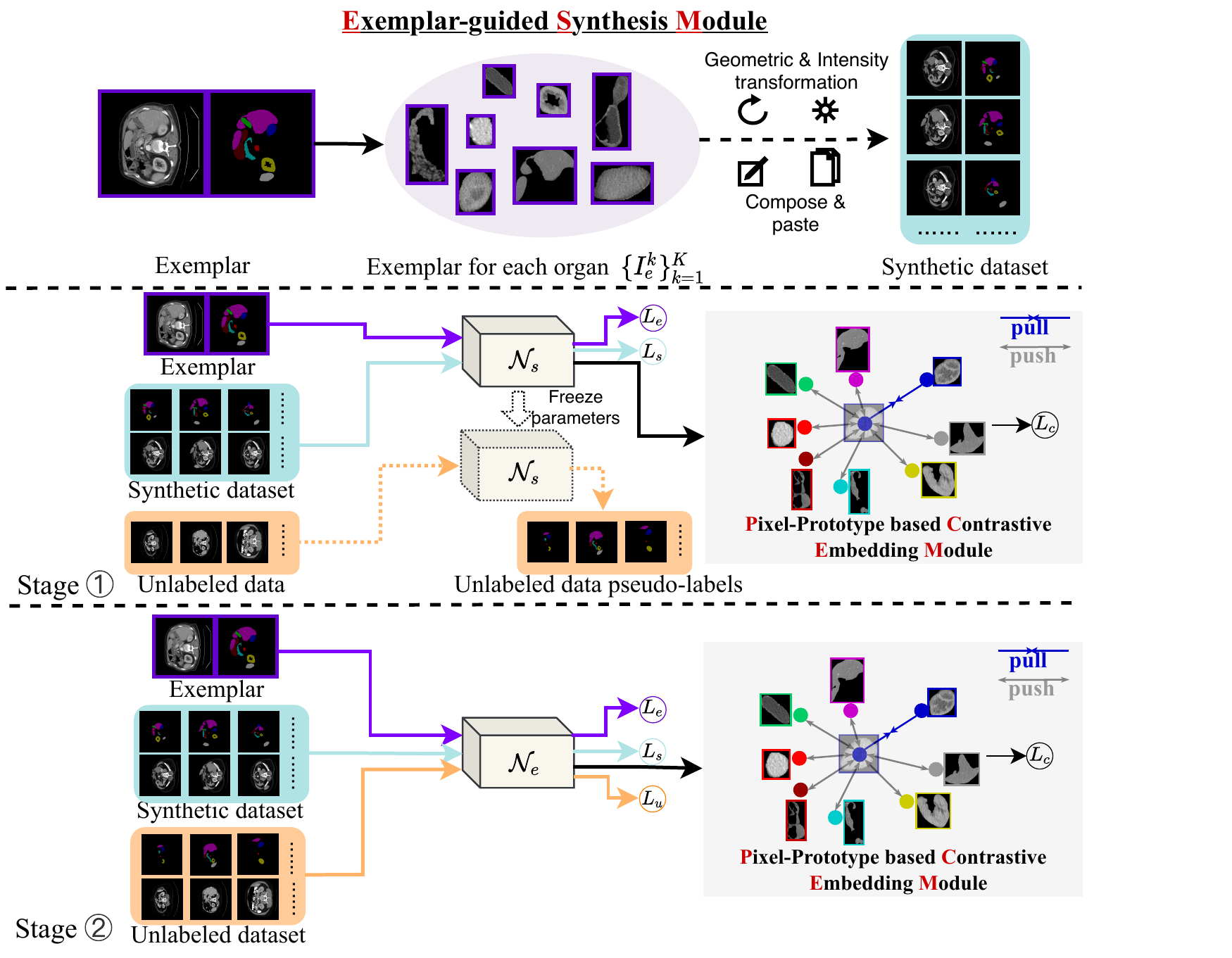}\\
\caption{
An overview of the proposed ELSNet framework.
First, ESM enriches the training data diversity.
In stage 1,	the segmentation network $\mathcal{N}_{s}$ is trained on the exemplar and synthetic data with PCEM and then used to generate the pseudo-labels of unlabeled data.
In stage 2, the exemplar, synthetic and unlabeled data are all used to train the segmentation network $\mathcal{N}_{e}$ with PCEM.
}
\label{fig2}
\end{figure*}

The overall architecture of the ELSNet is illustrated in Figure \ref{fig2}. 
We first create the synthetic dataset from the given exemplar by the exemplar-guided synthesis module, which enriches and diversifies the training set.
Moreover, we deploy a two-stage process for training the segmentation model by using the pixel-prototype based contrastive embedding module, which enhances the discriminative capacity of the base segmentation model and exploits the unlabeled data with predicted pseudo segmentation labels.
In the first stage, a synthetic segmentation network $\mathcal{N}_{s}$ is trained on the exemplar and the synthetic dataset, which is then used to generate pseudo-labels for unlabeled data.
In the second stage, an exemplar learning segmentation network $\mathcal{N}_{e}$ is trained on the exemplar, the synthetic dataset and the unlabeled dataset (with pseudo labels).
The two segmentation networks share the same structure,
consisting of an encoder $f_{encoder}: \mathbb{R}^{1*H*W} \to \mathbb{R}^{c*h*w}$ and a decoder $f_{decoder}: \mathbb{R}^{c*h*w} \to \mathbb{R}^{K*H*W}$, where
$c, h,$ and $w$ are the channel, height, and width of the embedding matrix, which is represented as $X=f_{encoder}(I)$. The decoder's output is used to produce the segmentation masks. 
Below we elaborate the two modules of the ELSNet framework and the training process.

\subsection{Exemplar-Guided Synthesis Module}\label{Exemplar-guided Sample Synthesis Module}
This module aims to synthesize the segmentation instance of each label category from the single exemplar into various backgrounds, thus creating a synthetic training dataset.
The diagram of this module is shown in Figure \ref{fig2}.
Given an exemplar, we first obtain the 
segmentation instance
of each organ according to their annotations, 
then
transform them to imitate the various 
appearances of the
organs 
in medical images, and finally paste the transformed organs onto background images.
We define a series of geometric transform operations and intensity transform operations as $\mathcal{T}_{g}$ and $\mathcal{T}_{i}$, respectively.
Then in principle, a synthetic sample $I_{s}$ and its label $Y_{s}$ can be generated by performing different transformations on $I_{e}$ and $Y_{e}$ as follows:
\begin{align}
(I_{s}, Y_{s}) = \Omega(\mathcal{T}_{g}(\mathcal{T}_{i}(I_{e})), \mathcal{T}_{g}(Y_{e}), \mathcal{T}_{g}(\mathcal{T}_{i}(I_{b})),
\label{transformation}
\end{align}
where $\Omega$ represents the proposed exemplar-guided synthesis operation that copies, transforms and pastes the exemplar onto the selected background image $I_{b}$.
We choose black images and images 
that do not contain any organs
as background images.
Meanwhile, the corresponding label $Y_{s}$ is generated based on the transformed $Y_{e}$.
By this means, the synthetic training dataset can be created by using Eq. (\ref{transformation}) with sufficient variations from one exemplar.

To better accommodate various transformations,
we propose to implement the $\Omega$ operation in a category-wise manner.
First, we segregate the exemplar $I_{e}$ into different categories of organs as follows: 
\begin{equation}
I_{e}^k = I_{e} \otimes Y_{e}^k,
\label{exemplar_organ}
\end{equation}
where $Y_{e}^k$ and $I^k_{e}$ indicate the mask and the exemplar 
instance for the \textit{kth} organ category, respectively.
Then, for the segmentation instance of each category from the single exemplar, geometric and intensity transformations are applied to imitate scale change, rotation, blur and intensity variations across the dataset.
Finally, we compose the transformed exemplar organs and paste them onto the background images, creating a synthetic sample.
Following this procedure, a synthetic dataset $\mathcal{D}_{S} = \{(I_{s}^b, Y_{s}^b)\}_{b=1}^{B}$ can be created, where $I_{s}^b\in \mathbb{R}^{1*H*W}$, $Y_{s}^b\in \{0, 1\}^{K*H*W}$ and $B$ is the number of synthetic samples.
With this synthetic dataset, a segmentation model can be trained without extra annotation effort. 
Moreover, as the basic features of all organs are present in the exemplar image, the synthetic dataset can overcome the limitations of previous works that rely on unrealistic virtual data 
\cite{dosovitskiy2015flownet,dosovitskiy2017carla}.

\subsection{Pixel-Prototype Based Contrastive Embedding Module}\label{Prototype Contrastive Module}
In this module, we calculate prototypes of different categories of organs and deploy a contrastive learning paradigm over the prototypes to improve the discriminability of the model.
Specifically,the organ prototypes are calculated based on the predicted masks from the decoder output. 
As previously stated, $X$ represents the embedding features of the input image, and we use $f_{decoder}$ to generate the predicted mask $\hat{Y}$ as follows:
\begin{equation}
\hat{Y}=argmax(softmax(f_{decoder}(X)))
,
\label{eq1}
\end{equation}
where $softmax$ denotes the class-wise 
softmax function and $\hat{Y}\in \mathbb{R}^{K*H*W}$ is the predicted label indicator matrix.
We resize the predicted mask $\hat{Y}\in\mathbb{R}^{K*H*W}$ to the same size as the embedding features $X$ via bilinear interpolation, which is denoted by $\hat{Y}_{x}\in\mathbb{R}^{K*h*w}$.
With $\hat{Y}_{x}$, we leverage global average pooling \cite{zhang2019canet,zhang2020sg} over the foreground to integrate the pixel
features belonging to the same category into a feature vector, which is seen as the prototype of the corresponding organ. 
The prototype for the $kth$ category, $v_k$, is computed as follows:
\begin{align}
\mathnormal{v}_{k}=\frac{\sum_{(i,j)}X^{(i,j)}\mathbbm{1}[\hat{Y}_{x}^{(k,i,j)}\neq0]}{\sum_{(i,j)}\mathbbm{1}[\hat{Y}_{x}^{(k,i,j)}\neq0]}
,
\label{SOFG}
\end{align}
where $(k,i,j)$ indicates the spatial location index of the $kth$ category and $\mathbbm{1}$ represents the indicator function.
The prototypes computed through the masked average pooling can
extract global object representations for the target organs.

To maximize the representation similarity of the same organ among different images while simultaneously minimizing the similarity of different organs, we propose to 
deploy a contrastive learning loss to learn discriminative embeddings. 
We perform the calculation in a batch of $N$ samples (\eg, a mini-batch). 
Specifically, for each category k and a prototype $v_k^n$ from the $nth$ image, 
we randomly select a prototype $v_k^m\in\{v_k^i, i\not= n | v_k^n\}$ from the other $N-1$ images in the current batch as a positive sample, and use the prototypes from all other categories of the $N-1$ images as negative samples.
Hence
the prototype-based contrastive loss is defined as follows:
\begin{equation}
L_{c}=-\sum_{n=1}^{N}\sum_{k=1}^{K}\log\frac{exp(v_{k}^{n} \cdot v_{k}^{m} / \tau)}{exp(v_{k}^{n} \cdot v_{k}^{m} / \tau) + \sum_{j \neq k}\sum_{i \neq n} exp(v_{k}^{n} \cdot v_{j}^{i} / \tau)}
,
\label{cont_loss}
\end{equation}
where $\tau$ is the temperature hyper-parameter;
$n$ is the index of the images.

This proposed module is designed with two main differences from previous self-supervised approaches.
First, 
our contrastive loss over the prototypes of various organs relies on the prediction mask rather than the entire image, which is directly related to the objective of the segmentation task.
Second, instead of augmenting the inputs to produce multiple copies, 
we perform contrastive learning across multiple images, aiming to capture
the semantic representation of the same organ that is invariant across different images.
Overall, by enforcing the embeddings of the same organ category 
to be similar in different images than that of different organ categories, 
PCEM can enhance the discriminability of the embedding learning. 

\subsection{Two-Stage Training}
The ELSNet is trained in two stages.
In the first stage, the exemplar $\mathcal{D}_{E}$ and the synthetic dataset $\mathcal{D}_{S}$ are used as training data to train the segmentation network $\mathcal{N}_{s}$ by minimizing the following joint loss function:
\begin{equation}
L_{s1} = L_{e}+\lambda_{s}L_{s}+\lambda_{c}L_{c}
\label{stage2loss}
\end{equation}
where $\lambda_s$ and $\lambda_c$ are trade-off hyperparameters, 
$L_{c}$ denotes 
the prototype-based contrastive loss defined in Eq.(\ref{cont_loss}),
$L_{e}$ and $L_{s}$ denote the segmentation losses 
computed from the exemplar and the synthetic dataset respectively,
such that 
$L_{e}=\mathcal{L}_{seg}(\hat{Y_{e}},Y_{e})$ and 
$L_{s}=\mathcal{L}_{seg}(\hat{Y_{s}},Y_{s})$. 
The segmentation loss $\mathcal{L}_{seg}$ is defined as follows:
\begin{equation}
\mathcal{L}_{seg}(\hat{Y},Y)=0.5*l_{ce}(\hat{Y},Y)+0.5*l_{dice}(\hat{Y},Y)
,
\label{eqseg}
\end{equation}
where $l_{ce}$ is the cross-entropy loss function and $l_{dice}$ is the Dice loss function;
$\hat{Y}$ denotes the predicted output of the segmentation network during the training stage.

After training the segmentation network $\mathcal{N}_{s}$, we use it  
to segment each image $I_u$ in the unlabeled set 
$\mathcal{D}_{U}$ and obtain its predicted segmentation mask $Y_{u}$
as the pseudo-labels, thereby constructing 
a pseudo-labeled set $\mathcal{D}_{U} = \{(I_{u}^t, Y_{u}^t)\}_{t=1}^{T}$.
Then in the second stage of training, 
we train our final segmentation network $\mathcal{N}_{e}$ by using all three sets of images, 
the exemplar $\mathcal{D}_{E}$, the synthetic dataset $\mathcal{D}_{S}$ and the 
pseudo-labeled $\mathcal{D}_{U}$ 
by minimizing the following joint loss function:
\begin{equation}
L_{s2} = 
	L_{e}+\lambda_{s}L_{s}+\lambda_{c}L_{c}+\lambda_{u}L_{u}
\label{stage3loss}
\end{equation}
where $\lambda_u$ is a trade-off hyperparameter, and the segmentation loss on the unlabeled data, 
$L_{u}=\mathcal{L}_{seg}(\hat{Y_{u}},Y_{u})$, is incorporated into training with their pseudo-labels. 
After the two-stage training, $\mathcal{N}_{e}$ can be used for inference on test images.

\section{Experiments}
\subsection{Experimental Setting}
\paragraph{{Implementation Details.}}
We adopt a U-Shape transformer-based structure \cite{wang2022mixed} as the basic structure of $\mathcal{N}_{s}$ and $\mathcal{N}_{e}$ in the proposed ELSNet.
The weights of the proposed ELSNet are randomly initialized.
The input image size is set to 224$\times$224 with random rotation and flipping.
Adam optimizes the proposed ELSNet with a weight decay of 0.0001 and a learning rate of 1e-4.
The batch size is set to 12, 
the patch size of the transformer \cite{wang2022mixed} is set to 16,
and the value $\tau$ is set to 0.07.
We divide all medical 3D volumes into individual images during the testing stage for inference \cite{wang2022mixed,chen2021transunet}. 
We randomly select a sample containing all categories in the training set as the exemplar.
In order to reduce the influence of fluctuations in the results, average results over five runs are reported for each experiment. 

\paragraph{{Datasets and Evaluation Metrics.}}
We evaluate the proposed framework on the Synapse dataset%
\footnote{https://www.synapse.org/\#!Synapse:syn3193805/wiki/217789} 
and the ACDC dataset%
\footnote{https://www.creatis.insa-lyon.fr/Challenge/acdc/}.
Synapse is a multi-label organ dataset containing 30 abdominal clinical CT cases with 2211 images, and we use 18 cases for training and 12 cases for testing \cite{chen2021transunet,wang2022mixed}.
ACDC is a cardiac MRI dataset that contains 100 cases from MRI scanners with 1300 images.
We used 70 cases for training, 20 for evaluation, and 10 for testing.
Following \cite{fu2020domain},
we evaluated the proposed framework based on two metrics, namely, the Dice Similarity Coefficient (DSC) and the 95\% Hausdorff Distance (HD95).

\subsection{Experimental Results} 
\subsubsection{Comparison Results} 
We compared the ELSNet with three state-of-the-art image segmentation methods under the same experimental setting on the ACDC and Synapse datasets: UNet \cite{ronneberger2015u}, MT-UNet \cite{wang2022mixed} and MLDS \cite{reiss2021every}. 
We re-implemented these methods and trained them under the same setting as the proposed ELSNet.
The comparison results on the two datasets are reported in Table \ref{tab:ACDC} and Table \ref{tab:Synapse}, respectively. 
On the ACDC dataset, the proposed ELSNet achieves considerable improvements, outperforming the second-best method, the semi-supervised MLDS, by 0.221 and 23.39 in terms of the class average DSC and HD95 results, respectively.
The Synapse dataset contains more organ categories with different sizes.
The foregrounds and background are more difficult to distinguish in the images of this dataset. 
Nevertheless, on the Synapse dataset, ELSNet again outperforms all three comparison methods and produces the best class average DSC result of 0.315 and the best class average HD95 result of 109.70. 
Moreover, ELSNet achieves the best results in almost all categories except {\em Spl}. 
The improvements are particularly large in the Aor and Gal categories. 
Overall, these results validate the efficacy of the proposed ELSNet for EL. 

%%%%%%%%%%%%%%%%%%%%%%%%%
\begin{table}[t!]\footnotesize
\renewcommand\arraystretch{1.2}
\caption{\label{tab:ACDC}
Comparison results on the ACDC dataset.
The class average results and the results for individual classes in terms of DSC and HD95 are reported.
}
\begin{center}
\setlength{\tabcolsep}{1.0mm}
\begin{tabular}{c|cccc|cccc}
\hline
Method&DSC.Avg$\uparrow$&RV&Myo&LV&HD95.Avg$\downarrow$&RV&Myo&LV\\
\hline
UNet\cite{ronneberger2015u}& 0.142 &0.140 &0.112&0.174&43.30&63.76&35.60&30.80\\
MT-UNet\cite{wang2022mixed}& 0.142&0.119&0.126&0.182&74.20&83.91&61.48&77.22\\
MLDS\cite{reiss2021every} &0.189&0.144&0.165&0.258&50.03&72.13&30.20&47.77\\
ELSNet
&\textbf{0.410}&\textbf{0.293}&\textbf{0.374}&\textbf{0.563}&\textbf{26.64}&\textbf{47.63}&\textbf{16.58}&\textbf{15.73}\\
\hline
\end{tabular}
\end{center}
\vskip -.15in
\end{table}
%%%%%%%%%%%%%%%%%%%%%%%%%
%%
\begin{table}[t]\footnotesize
\renewcommand\arraystretch{1.2}
\caption{\label{tab:Synapse} Comparison results on the Synapse dataset. 
The class average DSC and HD95 results and the DSC results for all individual classes are reported. 
}
\setlength{\tabcolsep}{.9mm}
\begin{center}
\begin{tabular}{c|c|ccccccccc}
\hline
\textbf{Method} &HD95$\downarrow$ & DSC$\uparrow$ & Aor & Gal&Kid(L)&Kid(R)&Liv&Pan&Spl&Sto\\
\hline
	UNet\cite{ronneberger2015u}&132.42&0.160&0.026&0.167&0.177&0.154&0.649&0.015&0.059&0.033\\
	MTUNet\cite{wang2022mixed} &154.60&0.112 &0.066&0.108&0.155&0.053&0.352&0.008&0.046&0.102 \\
	MLDS\cite{reiss2021every} &159.26&0.221&0.057&0.147&0.306&0.183&0.638&0.038&\textbf{0.306}&0.090\\
ELSNet
	&\textbf{109.70}&\textbf{0.315}&\textbf{0.319}&\textbf{0.372}&\textbf{0.381}&\textbf{0.219}&\textbf{0.784}&\textbf{0.067}&0.276&\textbf{0.104} \\
\hline
\end{tabular}
\end{center}
\vskip -.1in
\end{table}
%%%%%%%%%%%%%%%%%%%

\subsubsection{Qualitative Evaluation}
\begin{figure}[t]
\centering
  \includegraphics[width=0.77\linewidth,height=63mm]{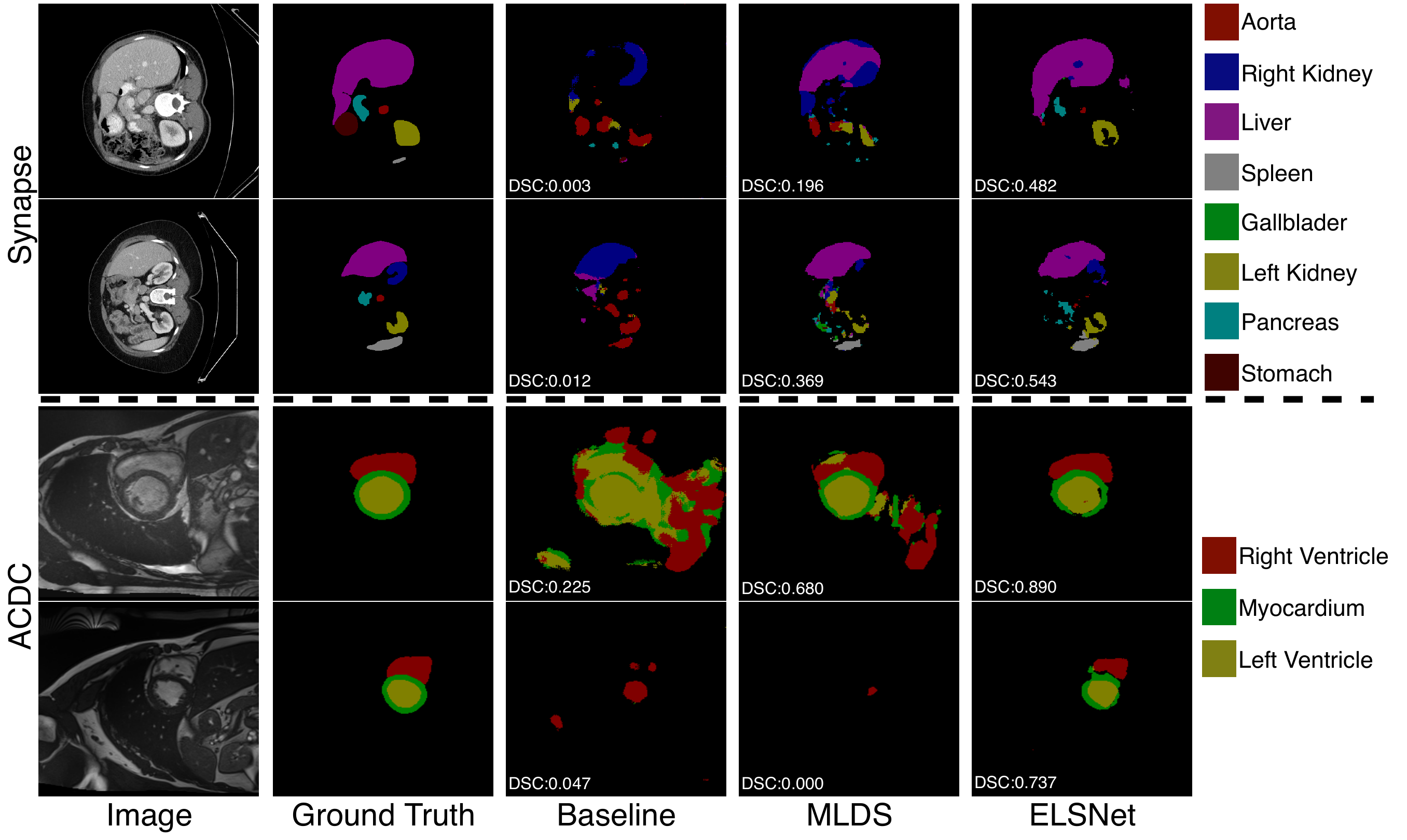}\\
\caption{
Visualized results obtained with the Baseline, 
	the MLDS and the proposed ELSNet on Synapse and ACDC datasets.
	The DSC value of each sample is also presented
}
\label{fig:vis}
\vskip -.1in	
\end{figure}

To further validate the segmentation performance of the proposed ELSNet, 
several visualized segmentation examples 
for Baseline, MLDS, and ELSNet are presented in Figure \ref{fig:vis}.
Baseline refers to the base segmentation network that is directly trained from the single annotated image.  
Typically, the tiny sizes of some organs can make segmentation very challenging, 
not to mention that there is only one annotation example.
However, we can see from Figure \ref{fig:vis} that compared with the existing state-of-the-art method, ELSNet can segment more accurately, even under conditions of deformation, edge ambiguity, shape complexity and background clutter shown in the examples.
%%%%%%%%%%%%%%%%%%%

%%%%%%%%%%%%%%%%%%%%%%%%%%%%%%%%%%%%%%
\subsubsection{Ablation Studies}
\begin{table}[t]
\vskip .2in
\renewcommand\arraystretch{1.3}
\caption{\label{tab:ablation_modules}
Ablation study of the proposed modules on the Synapse and ACDC datasets.
BS denotes 
the baseline.
+ESM denotes the variant that further includes the propsoed ESM. 
+ESM+PCEM\_S1 denotes the variant produced 
when both ESM and PCEM modules are used in stage 1.
+ESM+PCEM\_S1\&2 denotes the full approach. 
}
\begin{center}\footnotesize
\setlength{\tabcolsep}{1.1mm}
\begin{tabular}{c|cccc|cccc}
\hline
\multirow{2}*{Method}&\multicolumn{4}{c|}{Synapse}&\multicolumn{4}{c}{ACDC}\\
&DSC$\uparrow$&$\Delta$DSC&HD95$\downarrow$&$\Delta$HD95&DSC$\uparrow$&$\Delta$DSC&HD95$\downarrow$&$\Delta$HD95\\
\hline
BS&0.112&-&154.60&-&0.142&-&74.20&-\\
\hline
+ESM&0.234&\textcolor{red}{+0.122}&120.59&\textcolor{blue}{-34.01}&0.273&\textcolor{red}{+0.131}&38.35&\textcolor{blue}{-35.85}\\
\hline
+ESM+PCEM\_S1&0.264&\textcolor{red}{+0.152}&\textbf{101.00}&\textcolor{blue}{-53.60}&0.355&\textcolor{red}{+0.213}&40.80&\textcolor{blue}{-33.40}\\
\hline
+ESM+PCEM\_S1\&2&\textbf{0.315}&\textcolor{red}{+0.203}&109.70&\textcolor{blue}{-44.90}&\textbf{0.410}&\textcolor{red}{+0.268}&\textbf{26.64}&\textcolor{blue}{-47.56} \\
\hline
\end{tabular}
\end{center}
\vskip -.1in
\end{table}

\begin{table}[t]
\renewcommand\arraystretch{1.1}
\caption{\label{tab:ablation_loss}
Ablation results on ACDC.
{\bf Left side} presents the ablation results with different loss functions. 
	$L_{c}^{e+s}$ indicates that $L_c$ is used in Eq.(\ref{stage2loss}) for stage 1.
	$L_{c}^{e+s+u}$ indicates that $L_c$ is used in Eq.(\ref{stage3loss}) for stage 2.
{\bf Right side} presents the ablation results with different transformation strategies.	
Int.E and Int.B indicate intensity transformations are applied on Exemplar and Background images.
Similarly, Geo.E/Geo.B indicates geometry transformations.
}
\begin{center}
\setlength{\tabcolsep}{.8mm}
\resizebox{\textwidth}{!}{
\begin{tabular}{ccccc|cccc||cccc|cccc}
\hline
$L_{e}$&$L_{s}$&$L_{c}^{e+s}$&$L_{c}^{e+s+u}$&$L_{u}$&RV&Myo&LV&DSC.Avg$\uparrow$ &
Int.E&Int.B&Geo.E&Geo.B&RV&Myo&LV&DSC.Avg$\uparrow$\\
\hline
	$\surd$&-&-&-&-&0.119&0.126&0.182&0.142 &
-&-&-&-&0.136&0.159&0.232& 0.176\\
	$\surd$&$\surd$&-&-&-&0.192&0.265&0.362&0.273 &
$\surd$&-&$\surd$&-&0.255&0.317&0.362&0.311\\
	$\surd$&$\surd$&$\surd$&-&-&0.249&0.315&0.503&0.355 &
$\surd$&$\surd$&-&-&0.191&0.308&0.448&0.315\\
	$\surd$&-&$\surd$&$\surd$&$\surd$&0.263&0.323&0.488&0.359  &
-&-&$\surd$&$\surd$&0.210&0.356&0.451&0.339\\
	$\surd$&$\surd$&-&$\surd$&$\surd$&0.218&0.355&0.462&0.345 &
$\surd$&-&$\surd$&$\surd$&0.242&0.333&0.464&0.346\\
	$\surd$&$\surd$&$\surd$&-&$\surd$&0.290&0.354&0.529&0.390 &
$\surd$&$\surd$&$\surd$&-&0.272&0.334&0.502&0.370\\
	$\surd$&$\surd$&$\surd$&$\surd$&$\surd$&\textbf{0.293} &\textbf{0.374} &\textbf{0.563}&\textbf{0.410} &
$\surd$&$\surd$&$\surd$&$\surd$&\textbf{0.293}&\textbf{0.374}&\textbf{0.563}&\textbf{0.410}\\
\hline
\end{tabular} }
\end{center}
\vskip -.1in
\end{table}

\noindent\textbf{{Impact of the proposed modules.\ }}
We tested 
the empirical contributions of the proposed modules on the two datasets and the results are reported in Table \ref{tab:ablation_modules}.
Following the general experimental setup in \cite{reiss2021every,seibold2021reference}, BS denotes the baseline that uses only the exemplar as supervision, which achieves 0.112 and 0.142 in terms of DSC measure on the two datasets, Synapse and ACDC, respectively.
By adding
the proposed exemplar-guided synthesis module, ``+ESM'' substantially improves the performance 
to 0.234 and 0.273 in terms of DSC on the two datasets.
Such performance gains highlight the impact of the synthetic dataset produced by ESM, which enriches the diversity of the labeled samples and increases the generalization capability of the model.
By further including the proposed pixel-prototype based contrastive embedding module in stage 1, 
the results on the two datasets 
reach DSC values of 0.264 and 0.355 respectively. This demonstrates the impact of the PCEM module on enhancing the discriminative ability of the segmentation model.
Finally, the full model, "+ESM+PCEM\_S1\&2", produces the best results on both datasets by using all of the proposed modules in stage 1 and 2.
These results validate the impact of the proposed modules to the overall performance of the proposed ELSNet framework.
\\

\noindent\textbf{{Impact of different loss functions.\ }}
To demonstrate the effectiveness of the loss functions
involved in the training stages of ELSNet, 
we summarize the ablation results over multiple loss terms on the left side of Table \ref{tab:ablation_loss}.
The results in the second row show that creating the synthetic dataset with ESM ($L_s$) 
can significantly improve the DSC results to 0.273.
Using two modules, ESM ($L_s$) and PCEM ($L_c^{e+s}$), simultaneously to train the model in stage 1 can lead to further substantial improvements,
as shown in the third row.
Also, note that ignoring the synthetic dataset but still using the pseudo-labels of unlabeled data in stage 2 can 
obtain a result value of 0.359, which shows the unlabeled data is useful. 
When the contrastive loss $L_c$ is only used in either stage 1 or stage 2, 
the DSC results (in the sixth row and fifth row) degrade from the full model (the last row) 
that uses $L_c$ in both stages.
The full model with all the loss terms yields the best results.\\ 

\noindent\textbf{{Impact of the transformation strategies.\ }}
We tested different transformation strategies for synthesizing images in ESM 
by applying Geometric (Geo: scaling, rotation) and Intensity (Int: blur, intensity variations) transformations on the exemplar and background images.
The results of different variants are summarized and reported on the right side of Table \ref{tab:ablation_loss}. 
When none of the transformations is utilized, \ie, we paste the exemplar into different background images to build the synthetic dataset, minor improvement is achieved over the baseline that only uses the exemplar, as shown in the first row.
When both geometric and intensity transformations are applied to the exemplar, the average DSC performance is substantially improved to 0.311 (second row).
By applying the two types of transformations on both the exemplar and the background images, the performance of the full model improves by another 10\% or so to 0.410 (the bottom row). 
We also observe from the third and fourth rows that dropping either type of transformations (Geo or Int) can degrade the performance. 
These results suggest that both types of transformation are effective for generating useful synthetic data and the proposed ESM is reasonable. 
Moreover, dropping the intensity transformation on the background (fifth row) 
leads to larger performance degradation
than dropping the geometric transformation on the background (sixth row),
which indicates that the intensity diversity of the background is more important than the geometric diversity.

\section{Conclusions}
This paper introduced 
a new experimental scenario, Exemplar Learning, and proposed a novel framework, ELSNet, to learn segmentation models from only one annotated image.
ELSNet uses an exemplar-guided synthesis module (ESM) to enrich and diversify the training data by synthesizing annotated samples from the given exemplar, and uses a pixel-prototype based contrastive embedding module (PCEM) to increase the discriminative ability of the segmentation model by contrastive embedding learning.
A two-stage training process is deployed to exploit the unlabeled data via pseudo-labels. 
Experimental results demonstrate that the proposed framework is effective and outperforms existing segmentation methods under scenarios with very limited supervision information.

\bibliography{egbib}
\end{document}